\documentclass[prb,preprint]{revtex4-1} 

\usepackage{amsmath} 

\usepackage{graphicx,epstopdf}
\usepackage{epic}
\usepackage{amssymb}
\usepackage{lineno}
\usepackage{rotating}
\usepackage{amsmath}
\usepackage{afterpage}
\usepackage{epsfig}
\usepackage{multirow}
\usepackage{setspace}
\usepackage{longtable}

\begin{document}

\title{Decoding Pure Rotational Molecular Spectra for Asymmetric Molecules}


\author{S. A. Cooke}
\email{stephen.cooke@purchase.edu}
\author{P. Ohring}
\affiliation{School of Natural and Social Sciences, Purchase College SUNY, 735 Anderson Hill Road, Purchase, New York 10577, U.S.A.}

\date{\today}

\begin{abstract}
In this paper we demonstrate how asymmetric molecular rotational spectra may be introduced to students both ``pictorially" and with simple formulae.  It is shown that the interpretation of such spectra relies heavily upon pattern recognition.  The presentation of some common spectral patterns in near-prolate asymmetric rotational spectra provides a means by which spectral assignment, and approximate rotational constant determination, may be usefully explored in the physics and chemistry classrooms.  To aid in this endeavor we have created a supporting, free, web page and mobile web page.
\end{abstract}

\maketitle

\section{Introduction}

Microwave spectroscopy is capable of providing unique insights into the electronic structure and potential energy surfaces of molecules.  It has found considerable use in the field of physics and astronomy, providing an invaluable method to identify chemical species present in the interstellar medium.  Microwave spectroscopy is a mature discipline and several recent review articles are available on the subject\cite{nwalker,grabcam1,grabcam2}.  Yet recent technological advances\cite{cp}, together with lower instrument costs mean that microwave spectroscopy is more accesible to undergraduate and graduate students than it has ever been before\cite{marshall,jennifer,sean,steve,gordon,saci}.  Furthermore, the technology developments now allow broadband rotational spectra to be collected in a few hours, or less, meaning that students can perform experiments in a timely fashion, i.e. an undergraduate physical chemistry lab.  This being so, the broadband nature of the data collected means that students are still faced with a complex spectrum that may not be as easily interpreted as many classroom NMR spectra or low resolution IR spectra.

The rigid rotor, as a model for a rotating linear molecule, is a subject covered in elementary quantum mechanics classes.  It is unusual for classes to consider molecular asymmetry, because this is often viewed as being too complex a problem.  It is the goal of this work to make this important task easier.

All types of spectroscopy require that the observed transitions, or peaks, be identified.  For many types of spectroscopy spectral peak identification is associative.  For example associating a specific bond stretch with a peak in an IR spectrum, or associating a methyl group with a group of peaks in an NMR spectrum.  However, in many high resolution spectra, the task of transition assignment requires identifying the upper and lower quantum states invovled. 
The spectra produced from microwave frequency studies are usually associated with the quantized rotational motions of molecules, and the task of spectral peak identification involves the latter, i.e. the assignment of quantum numbers for the upper and lower states.

A rotational spectrum, like most spectra, should be viewed as a puzzle.  These puzzles contain patterns, often complex in nature, that once unravelled lead to easily tractable information regarding the geometry and conformation of the subject molecule.  
In order for spectral patterns to be unravelled the spectroscopist needs to know the ``rules" of the game, i.e. the underlying theory governing the asymmetric quantum rotor.  For any given molecule these rules may be arrived at from an appropriately formed Hamiltonian operator, solving for the eigenvalues, and then either deriving or simply applying the appropriate selection rules.  This could be called a bottom-up approach.  But there is an alternative where, first, puzzle solving skills are invoked and spectral patterns identified.  After this point the appropriate molecular properties may be introduced and discussed.  This may be considered a top-down approach.  

To this end several, powerful, software visualization tools have been developed to help with the quantum number assignment problem in pure rotational spectroscopy\cite{kis,prospe,pgopher,caaars,jb95}.  In order to compliment these tools the provision of some common spectral patterns would be useful for the beginning student. 

The aim of this paper is to present some common patterns that occur in rotational spectra, together with their approximate dependence on the magnitudes of the rotational constants, so that a top down approach is more available to interested parties.  None of the theory introduced here is new (its over 70 years old\cite{wang,ray,king}), however, the method of presentation goes beyond that found in other sources.  Through this article it is hoped that students may be more easily introduced to rotational spectra, and that rotational spectroscopy may be more fully included in the chemistry curriculum.  

\section{Background}

In order for spectral patterns to be identified, and in order to produce a self contained article, three concepts need to be introduced.  These are (i) molecular rotational constants, (ii) the appropriate quantum numbers to use in transition assignment, and (iii) an overview of the selection rules with particular reference to how these selection rules relate to the geometry of the molecule.  These are introduced in only a very cursory manner as they are treated thoroughly by several excellent monographs\cite{strandberg,townes,allen,wollrab,flygare,kroto,gordy}.

\subsection{Molecular Rotational Constants}
The geometry of any molecule may be reduced to three rotational constants, $A$, $B$, and $C$, which are inversely proportional to the moments of inertia about the three principal axes, $a$, $b$, and $c$.  By convention, the $abc$ axis system is orientated such that the $A$ rotational constant is largest in magnitude, and $C$ is smallest.  Accordingly, an important skill is to be able to locate the $a$ principal axis in such a way that it runs through, or close to, as much atomic mass as possible, followed by the $c$ principal axis such that it is perpendicular to the $a$ axis and runs through as little atomic mass as possible.  For examples of this axis system Figure 1 displays three different molecules in their principal axes.  The three molecules are thioxoacetaldehyde\cite{thio} (top), hexafluoroisobutene\cite{hexa} (center), and 1,1,1,3,3,3-hexafluoropropane\cite{ctype} (bottom).

\begin{figure}[h!]
\centering
\includegraphics[width=4.0 in]{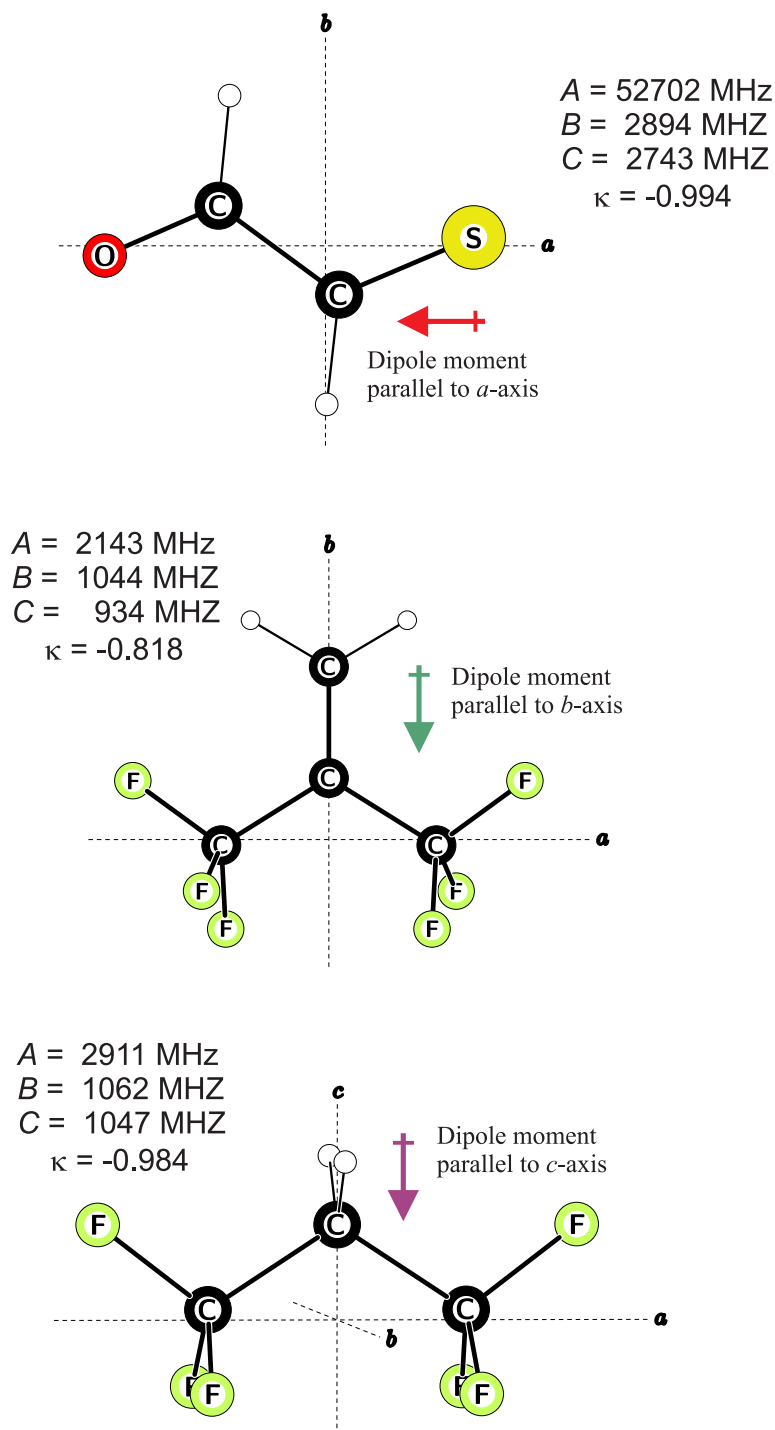}
\caption{An example of three molecules in the $abc$ principal axes system.  Notice how, in each case, the $a$-axis is located such that it is as close to as much atomic mass as possible.  As evidenced by the inset $\kappa$ values and rotational constants, all three molecules may be considered near-prolate asymmetric rotors.  The direction of the dipole components for each molecule are shown.  The lines between atoms indicate connectivity and not bond order. The three molecules are thioxoacetaldehyde (top), hexafluoroisobutene (center), and 1,1,1,3,3,3-hexafluoropropane (bottom). }
\label{fig:dipole}
\end{figure}

The relationship between the magnitudes of the rotational constants, i.e. the shape of the molecule, governs the gross appearance of a molecules rotational spectra.  Ray's asymmetry parameter\cite{ray}, $\kappa$ = (2$B$-$A$-$C$)/($A$-$C$), provides a quantitative measure of how far a molecule is from any symmetry relations.  In the symmetric prolate limit where $A > B = C$, then $\kappa$ = -1, whereas in the symmetric oblate limit where $A = B > C$, $\kappa$ = +1.  Figure 2 displays the moment of inertia ellipsoids for the prolate, perfect asymmetric, and oblate tops.  The ellipsoids provide a visual cue of the relationships between the moments of inertia, and help the student visualize molecular rotation for an asymmetric top by imagining the ellipsoids spinning on a surface. Naturally, most molecules are asymmetric and, as may be observed in Figure 2, possess moment of inertia relationships that lie somewhere between the oblate and prolate limits.  For many molecules $\kappa$ is close to -1 and they are then described as near-prolate asymmetric tops.  This is a very common molecular shape and this paper focuses mainly on this type of rotational spectrum.   

\begin{figure}[h!]
\centering
\includegraphics[width=4.0 in]{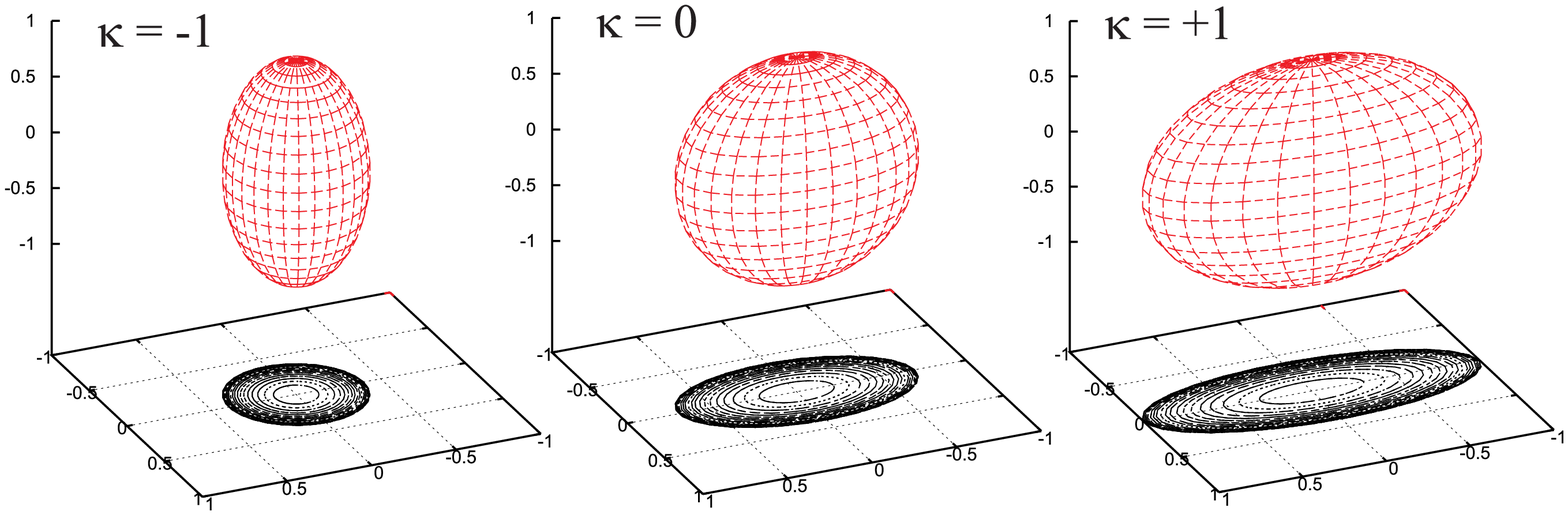}
\caption{The moment of inertia ellipsoids for symmetric prolate (left), asymmetric (center), and symmetric oblate (right) molecular rotors.}
\label{fig:elipsoid}
\end{figure}

\subsection{Rotational Quantum Numbers for Asymmetric Molecules}

The rotational quantum numbers for asymmetric tops are denoted as $J_{K_{-1}K_1}$.  The rotational quantum number $J$ is familiar from the linear rigid rotor, however the $K_{-1}$ and $K_1$ labels (not quantum numbers) require brief explanation.  Consider a symmetric top, either oblate or prolate. A second quantum number, $K$, is required in the energy expressions for these molecules.  This quantum number pertains to the quantized rotation about the molecules symmetry axis. The quantum number $K$ can take values of $J$ through to -$J$ in integer steps.  However, only a $K^2$ term appears in the energy expression and so, excepting the $K$ = 0 levels, all K levels are doubly degenerate.  Now, for a prolate top, the energy levels for a specific quantum state, $J_K$, increase in energy from $J_{0}$ up to $J_J$.  In stark contrast to this the oblate top energy levels, $J_K$, increase in energy from $J_{J}$ up to $J_0$.

For the asymmetric top there is no symmetry axis.  This means $K$ is not a good quantum number for an asymmetric molecule. However, each energy level for the asymmetric top may be linked to (i) a $K$ value in the prolate limit, this becomes the $K_{-1}$ label, and also (ii) a $K$ value in the oblate limit, which becomes the $K_1$ label.  Essentially, we may imagine that if an asymmetric top in a $J_{K_{-1}K_1}$ state were to ``morph" into a prolate top, then the quantum state would be $J_{K_{-1}}$.  If, on the other hand, the asymmetric $J_{K_{-1}K_1}$ state were to ``morph" into an oblate top the quantum state would be $J_{K_1}$.

\subsection{Selection Rules}

The gross selection rule for rotational spectroscopy is that the molecule must have a dipole moment.  In the case of an asymmetric molecule one must consider the  components of the dipole along either the $a$ and/or $b$ and/or $c$ principal axes.  A key skill requires that electronegative/electropositive elements, or groups, be identified for the subject molecule within the $abc$ axis system.  In this way one can tell $a$ $priori$ along which axis or axes a dipole component will be non-zero.  Figure 1 shows three examples of molecules with either $a-$, $b-$, or $c-$dipole components.

A molecular rotation about the $a$ principal axis places different symmetry requirements on the relation between upper and lower wavefunctions in a rotational transition compared to a rotation about either the $b$ or $c$ axes.  These requirements are manifest in the relation between the upper and lower quantum states $K_{-1}K_1$ labels.  It is the parity of the labels, i.e. even ($e$) or odd ($o$), that we concern ourselves with.  In short, for a molecular rotation about the $a$-axis, along which there is a dipole component, a change in parity is only allowed for the $K_1$ label. This is referred to as an $a$-type transition.  For a $c$-type transition, a change in parity is only allowed in the $K_{-1}$ label.  For a $b$-type transition, a change in parity must occur in the $K_{-1}$ and $K_1$ labels simultaneously.

In regards to the changes permitted in the rotational quantum number $J$ it may be shown that $\Delta J = -1,0,+1$. Here $\Delta J$ means $J$(upper state, higher energy) - $J$(lower state, lower energy).  Just as is the case for vibration-rotation spectroscopy, transitions of the type $\Delta J = -1$ are referred to as $P$-branch transitions, and by extension $\Delta J = 0$ is the $Q$-branch, and $\Delta J = +1$ is the $R$-branch.

Lastly, a shorthand notation for any transition, or group of transitions, is commonly used.  By example, an $^aR_{0,1}$ transition is an $a$-type, $R$-branch transition in which the value of $K_{-1}$ does not change between the lower and upper states, and the $K_1$ label increases by 1 between lower and upper states.   

\section{A Note on Centrifugal Distortion}

In this work we have assumed the rotating molecule to be rigid.  Real molecules do, however, undergo distortions in their average nuclear positions as rotation occurs.  This is referred to as centrifugal distortion and is accounted for in a descriptive set of molecular parameters by several centrifugal distortion constants\cite{watson}.  The magnitudes of these constants are, crudely, inversely proportional to the square of the molecular mass. The centrifugal distortion constants are many orders of magnitude smaller than the rotational constants.  The effect of centrifugal distortion on the observed spectra is to change the frequency of a given rigid rotational transition.  The effects are generally minimal, i.e. a few kHz or so for low $J$ transitions, but can become significant, i.e. many MHz, for high $J$ transitions.  They have been ignored in this work because at low $J$, i.e. $J$ less than, say 10, centrifugal distortion does not effect the patterns of rotational transitions observed.  However, with modern microwave spectrometers, the precision of measurement of rotational transitions is such that the effects of centrifugal distortion are routinely observed and must be accounted for in the Hamiltonian operator.  

\section{Rotational Spectra}

In the following we have selected some common transition series observed in rotational spectra for asymmetric molecules.  The transitions are discussed using the notation $J'_{K_{-1}K_1} \leftarrow J''_{K_{-1}K_1}$, where the double prime indicates the lower state.

We describe them pictorially and by the provision of simple formula.  It should once again be noted that effects of centrifugal distortion have been neglected.  The spectra displayed in Figures 3 through 8 were calculated exactly using the rotational constants provided in each figure using a complete Hamiltonian matrix diagnolization routine\cite{pick1,pick2}. The calculated spectra relate to molecules at rotational temperatures of 3 K which is a common temperature for operating a modern microwave spectrometer. The simple formulae provided have been determined through consideration of both the closed algebraic expressions for the lowest $J$ energy levels\cite{townes} and also through $a$ $posteriori$ examination of exact spectra.  A useful compilation of empirical relations for the frequencies of rotational spectral transitions is provided by Gordy and Cook\cite{gordy}.  However, to the best of our knowledge, many of the formula provided here appear in print for the first time.  

In regards to strategy, assignment should begin by examining the entire rotational spectrum and looking for repeating patterns.  If patterns can be located they may be compared to the portions of spectra displayed in Figures 3 through 8.  Once patterns have been identified the empirical relations provided below, and summarized in Table 1, allow initial approximations of the rotational constants $A$, $B$, and $C$. With initial estimates of the rotational constants further patterns may be searched for and the process repeated in a bootstrap fashion. 

\clearpage
\begin{singlespace}
\begin{table}         

\caption{Summary of Transition Patterns and Features}
\label{tbl:rotnrg}

\begin{tabular}{llp{5cm}}
\hline

\multicolumn{3}{c}{\underline{$R$-branch, $J'$ = $J''+1$}}\\

Transition type & Frequency & Notes \\ \hline
&&\\

{\underline{$a$-type}} &&\\
          & ($B+C$)($J''+1$) & For each $J' \leftarrow J''$, with $J'' > 0$, transitions are spread over ($B-C$)($J''+1$) \\

{\underline{$b$-type}} &&\\
&&\\
$J'_{1,J'} \leftarrow J''_{0,J''}$ & $A + C + 2CJ''$& \\
&&\\
$J'_{0,J'} \leftarrow J''_{1,J''}$ & ($3B+2C-A$) + $2B$($J''-1$)   & \\
&&\\

$J'_{2,J'-1} \leftarrow J''_{1,J''-1}$ & \multirow{2}{*}{($3A+C+\frac{(B-C)}{2}$)+($B+C$)($J''-1$)} & \multirow{2}{*}{Doublet split by $\frac{(B-C)}{2}J'J''$}\\

$J'_{2,J'-2} \leftarrow J''_{1,J''}$   &                                     & \\

&&\\
$J'_{3,J'-2} \leftarrow J''_{2,J''-2}$ & \multirow{2}{*}{($5A+C+\frac{(B-C)}{2}$)+($B+C$)($J''-2$)} & \multirow{2}{*}{Very closely spaced doublets}\\
$J'_{3,J'-3} \leftarrow J''_{2,J''-1}$ &                                     & \\

&&\\
{\underline{$c$-type}} &&\\
&&\\
$J'_{1,J'-1} \leftarrow J''_{0,J''}$ & $A + B + 2BJ''$& \\

&&\\
\multicolumn{3}{c}{\underline{$Q$-branch, $J'$ = $J''$}}\\

Transition type & Frequency & Notes \\ \hline
&&\\

{\underline{$a$-type}} &&\\
&&\\
$J_{1,J-1} \leftarrow J_{1,J}$ & $J$ = 2: $\nu$ = 3($B-C$) & $J>2$: \\
                                                            && $\nu (J) = \nu (J-1)+(B-C)J$ \\
&&\\
$J_{J,1} \leftarrow J_{J-2,2}$   & $2(A-2(B+C))(J-1)$ & Groups of same $K_{-1}$\\

&&\\

{\underline{$b$-type}} &&\\
&&\\
$J_{2,J-2} \leftarrow J_{1,J-1}$ & $J$ = 2: $\nu$ = 3($A-B$) & $J>2$: \\
                                                            && $\nu (J) = \nu (J-1)-\frac{(B-C)}{2}J$ \\
&&\\
$J_{2,J-1} \leftarrow J_{1,J}$   & $J$ = 2: $\nu$ = 3($A-C$) & $J>2$: \\
                                                            && $\nu (J) = \nu (J-1)+\frac{(B-C)}{2}J$ \\
&&\\
\hline
\end{tabular}
\end{table}
\end{singlespace}
\clearpage

\subsection{$R$-branch transitions}

\subsubsection{$a$-types}

These transitions are perhaps the easiest to locate in a broadband spectrum.  The ease with which they may be located is based on their regular harmonic pattern.  They occur in groups centered at
approximately ($B+C$)($J''$+1), i.e. every $B+C$ MHz a similar looking group is observable. Each group spans ($B-C$)($J''$+1). A sample of one of these groups is shown in Figure 3 for three different asymmetries.  
It is observed that the strongest transition in this group, the
$J'_{0,J'} \leftarrow J''_{0,J''}$ transition, moves to lower frequency, away from the center of the group, as $\kappa$ moves further away from -1.  

\begin{figure}[h!]
\centering
\includegraphics[width=4.0 in]{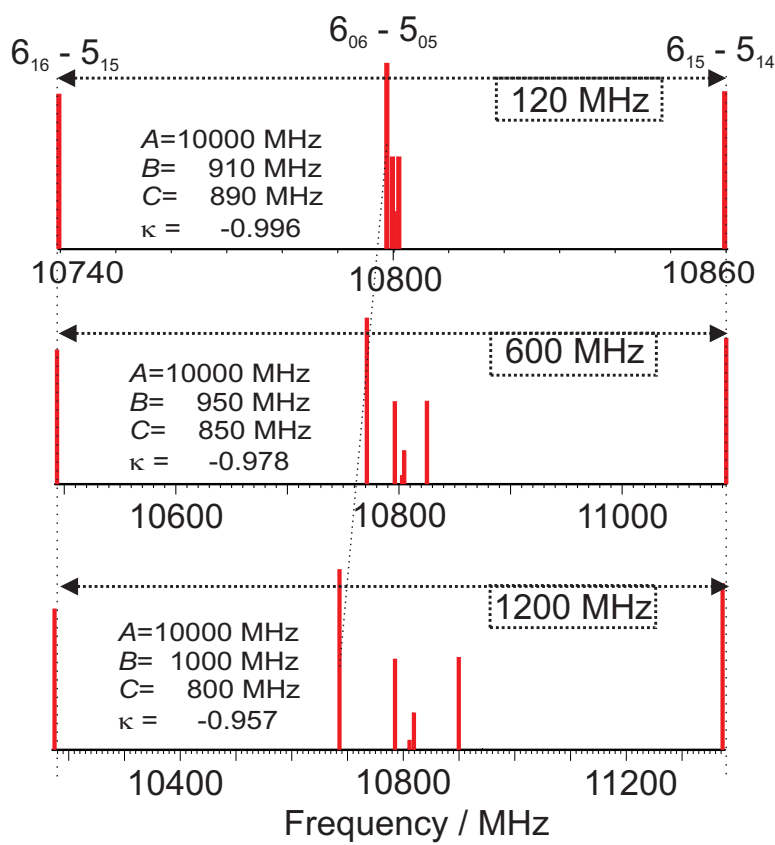}
\caption{An example of a common pattern for $a$-type $R$-branch transitions.  The three patterns show the effects of increasing asymmetry as the figure is viewed from top to bottom.  The center of the group is located at ($B+C$)($J''+1$), the spread of each group is equal to ($B-C$)($J''+1$).  The values of rotational constants used to generate the spectrum are shown.}
\label{fig:atyper1}
\end{figure}

\begin{figure}[h!]
\centering
\includegraphics[width=6.0 in]{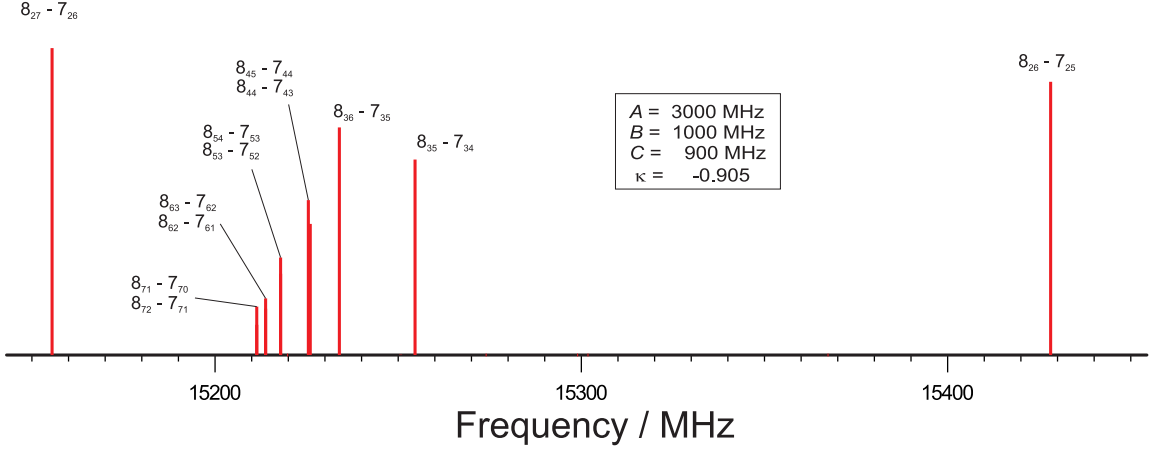}
\caption{A second example of a common pattern for $a$-type $R$-branch transitions in which the $K_{-1}$ label takes values larger than 2.  The transitions appear as doublets starting with the highest $K_{-1}$ doublet located at approximately ($B+C$)($J''+1$) with the successive lower $K_{-1}$ doublets occuring at higher frequency and increased splitting.}
\label{fig:atyper2}
\end{figure}

Within each group of transitions, and when the $A$ rotational constant is of modest magnitude compared to $B$ and $C$, transitions involving high values of the $K_{-1}$ label are observable as doublets in a characteristic pattern.  The pattern contains unresolved doublets, involving the highest $K_{-1}$ values, emerging from approximately ($B+C$)($J''+1)$.  The pattern progresses to higher frequencies, with doublets appearing with increased splittings as $K_{-1}$ decreases in value. An example is shown in Figure 4, and the pattern is also visible in Figure 3.  For low $J$ transitions, i.e. $J''<5$, the splitting in the $K_{-1}$=2 doublet is crudely given by $\left[ (B-C)^2 / A \right] (J''+1)$.  For higher $J$ transitions this estimate is approximately one order of magnitude too small.

\subsubsection{$b$-types}

\noindent {\bf{1. $J'_{1,J'} \leftarrow J''_{0,J''}$}}: The lowest frequency member of this group of transitions, i.e. the $1_{11} \leftarrow 0_{00}$ transition, occurs at $A$+$C$.  Transitions in this family then progress to higher frequencies spaced by 2$C$, giving a general formula of $A$+$C$+2$CJ''$.  An example of the progression is shown in Figure 5.

\begin{figure}[h!]
\centering
\includegraphics[width=6.0 in]{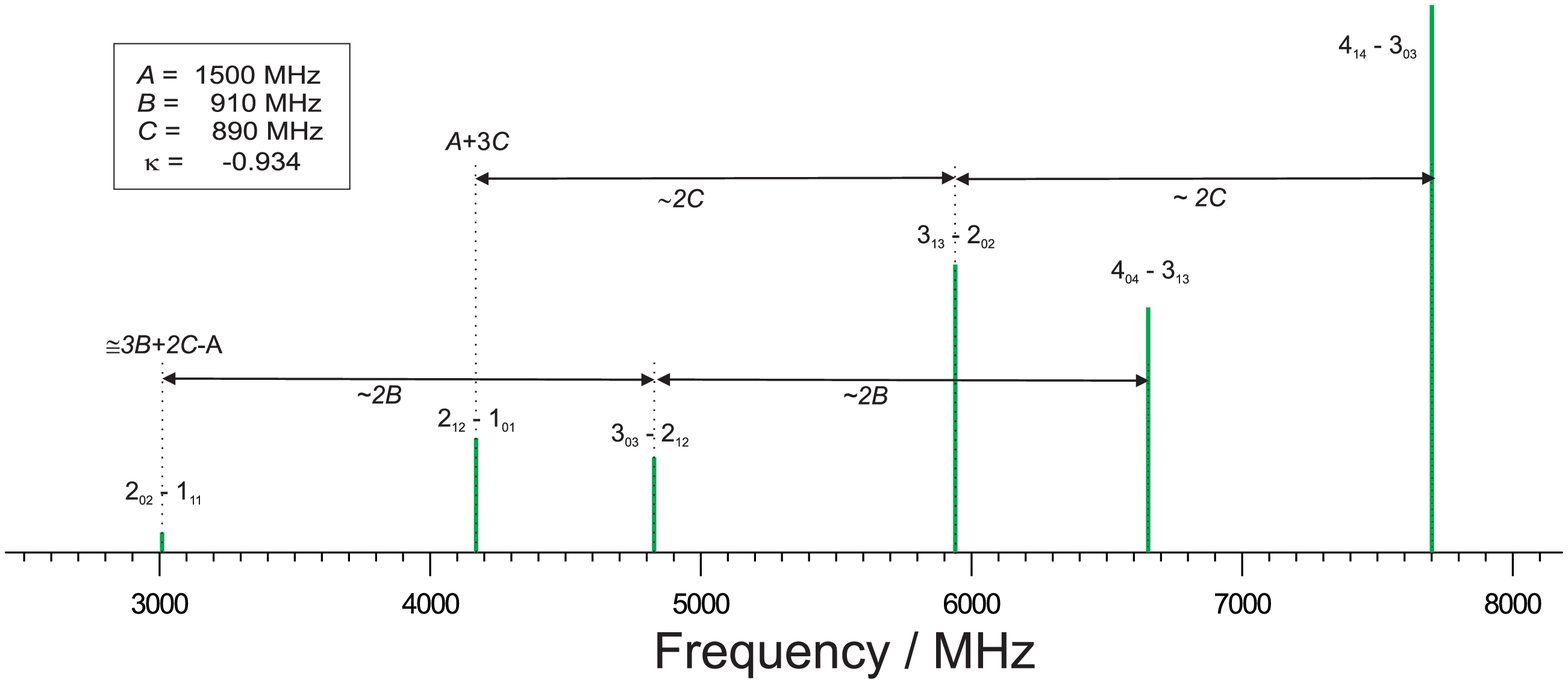}
\caption{Common $b$-type $R$-branch transition patterns.  Two sequences are shown, firstly the $J'_{1,J'} \leftarrow J''_{0,J''}$ transitions with approximate general formula $A+C+2CJ''$, and secondly the $J'_{0,J'} \leftarrow J''_{1,J''}$ transitions with approximate general formula $3B+2C-A+2B(J''-1)$.}
\label{fig:btyper1}
\end{figure}

\vspace{3mm}

\noindent {\bf{2. $J'_{0,J'} \leftarrow J''_{1,J''}$}}: The lowest frequency member of this group of transitions, i.e. the $2_{02} \leftarrow 1_{11}$ transition, occurs at $3B+2C-A$.  Transitions in this family progress to higher frequencies spaced by 2$B$.  This results in a general formula of $3B+2C-A$+2$B$($J''-1$). An example of the progression is shown in Figure 5.

\vspace{3mm}

\noindent {\bf{3. $J'_{2,J'-1} \leftarrow J''_{1,J''-1}$ and $J'_{2,J'-2} \leftarrow J''_{1,J''}$}}:  For a given $J''$ these two transition types appear as characteristic doublets within the rotational spectrum.  The lowest pair of transitions allowed are the $2_{21} \leftarrow 1_{10}$ and the $2_{20} \leftarrow 1_{11}$ transitions. The center frequency of the doublets in this family of transitions is given by $(3A+C+\frac{(B-C)}{2})+(B+C)(J''-1)$.  The splitting between the doublets is given by approximately $\frac{(B-C)}{2} J'J''$, i.e. half of $B-C$, multiplied by the product of the upper and lower $J$ values. Given the dependence of the frequency on 3$A$, these transitions often fall at higher frequencies than the previously listed transitions. For example, for the CH$_3$CHO the $2_{21} \leftarrow 1_{10}$ transition falls at approximately 179 GHz. An example of the progression is shown in Figure 6.

\begin{figure}[h!]
\centering
\includegraphics[width=6.0 in]{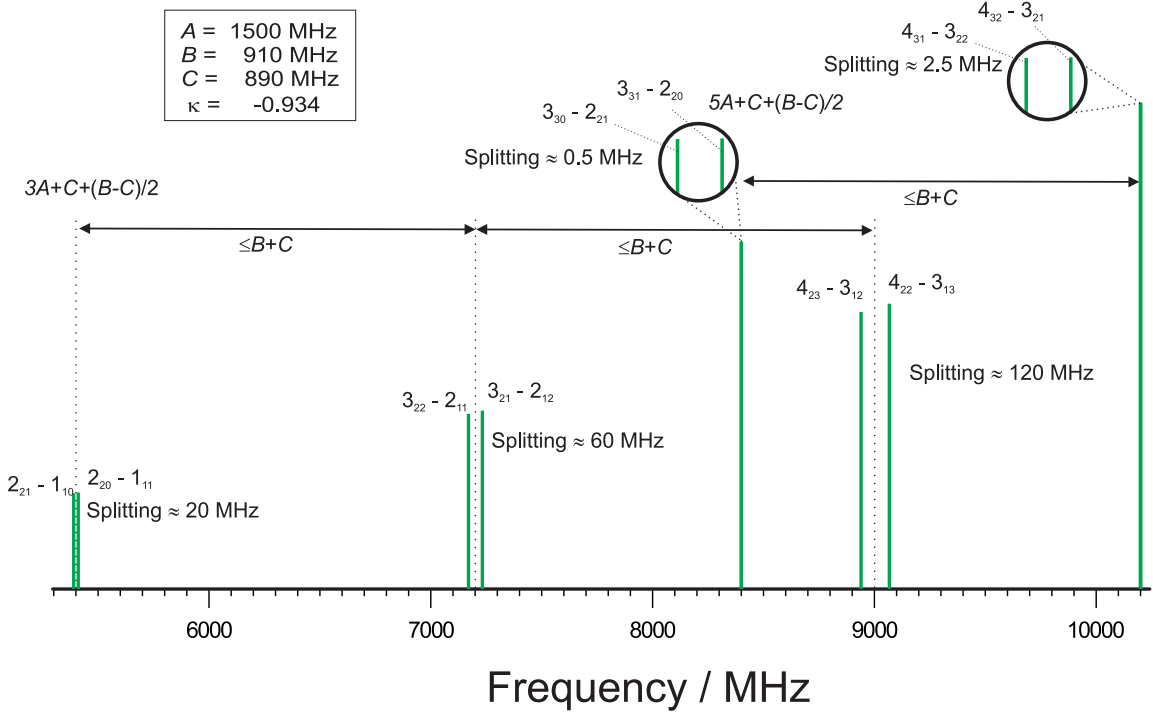}
\caption{Higher $K_{-1}$ $b$-type $R$-branch transition patterns.  Two sequences are shown, the text should be consulted for further description.  These doublets are often useful in making quantum number assignments.}
\label{fig:btyper2}
\end{figure}

\vspace{3mm}

\noindent {\bf{4. $J'_{3,J'-2} \leftarrow J''_{2,J''-2}$ and $J'_{3,J'-3} \leftarrow J''_{2,J''-1}$}}: For a given $J''$ these two transitions appear as very closely spaced doublets. For values of $\kappa$ approaching -1, where $B-C$ is small, the doublets are often not resolvable.  The center frequency for the doublets appearing in this family is given by $(5A+C+\frac{(B-C)}{2})+(B+C)(J''-1)$.  As was the case for the previous set of transitions, these doublets often fall at high frequencies, owing to their dependence on $5A$.  An example of the progression is shown in Figure 6.

\subsubsection{$c$-types}

\noindent {\bf{1. $J'_{1,J'-1} \leftarrow J''_{0,J''}$}}: The lowest frequency member of this group of transitions, i.e. the $1_{10} \leftarrow 0_{00}$ transition, occurs at $A$+$B$.  Transitions in this family then progress to higher frequencies spaced by 2$B$, giving a general formula of $A$+$B$+2$BJ''$.  An example of the progression is shown in Figure 7. We note in passing that it is not uncommon for $c$-type transitions and $b$-type transitions of the type listed above to occur in similar patterns.  Caution should be made to ensure that the non-zero dipole components of the molecule under study are identified prior to analysis (see Figure 1).

\begin{figure}[h!]
\centering
\includegraphics[width=6.0 in]{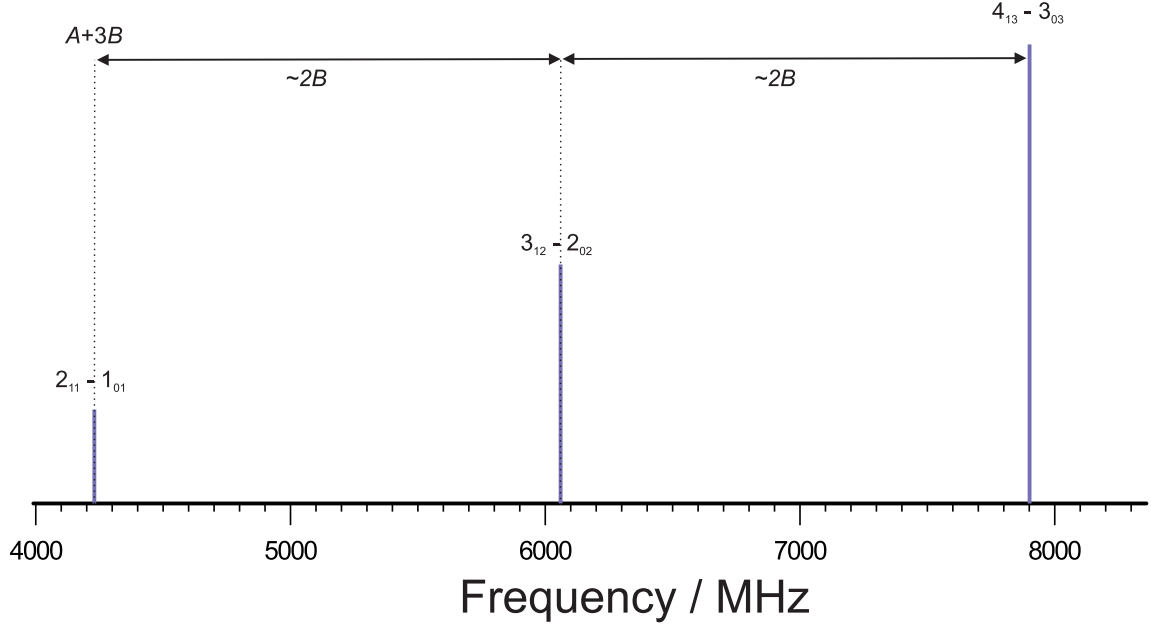}
\caption{A common $c$-type $R$-branch transition pattern.  The $J'_{1,J'-1} \leftarrow J''_{0,J''}$ transitions have an approximate general formula of $A+B+2BJ''$.}
\label{fig:ctyper}
\end{figure}

\subsection{$Q$-branch transitions}

\subsubsection{$a$-types}

\noindent {\bf{1. $J_{1,J-1} \leftarrow J_{1,J}$}}: The lowest frequency of these transitions, with $J$=2, occurs at 3$(B-C)$.  For $J > 2$ the transition frequencies increase according to the recursion formula $\nu (J) = \nu (J-1)+(B-C)J$.  For near prolate cases $B-C$ is a small number and accordingly only very high $J$ transitions would fall in observable regions where they will likely be low in intensity.

\noindent {\bf{2. $J_{J,1} \leftarrow J_{J-2,2}$}}: These transitions are found at $2(A-2(B+C))(J-1)$.  The transitions are found as closely spaced groups where the $J$ level changes, the $K_{-1}$ does not, and the $K_1$ levels increases by 1. That is, the $3_{31} \leftarrow 3_{12}$ transition will be found close by (a few tens of MHz) to the $4_{32} \leftarrow 4_{13}$ transition and so on.

\subsubsection{$b$-types}
These transitions produce very distinct patterns in rotational spectra.  An example is shown in Figure 8.  Transitions of the type $J_{2,J-2} \leftarrow J_{1,J-1}$ ``appear", i.e. when $J$ =2,  at at frequency of 3$(A-B)$.  The transitions then ``fan" out to lower frequencies with increasing $J$ according to the recursion formula $\nu (J) = \nu (J-1) - \frac{(B-C)}{2} J$. For near prolate molecules, the next group of transitions, for which $K_{-1}$ = 3, occur at approximately $A-\frac{(B-C)}{2}$ higher in frequency.

\begin{figure}[h!]
\centering
\includegraphics[width=6.0 in]{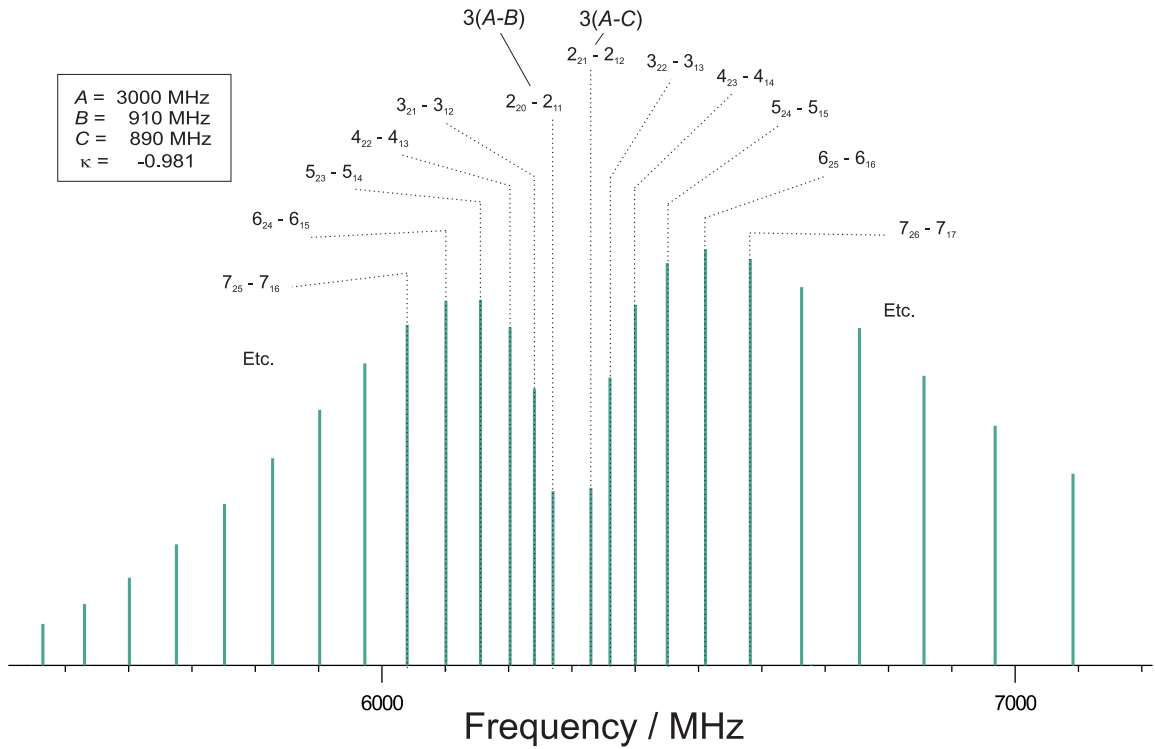}
\caption{An example of a $b$-type $Q$-branch progression of transitions.}
\label{fig:btypeq}
\end{figure}

Related transitions of the type $J_{2,J-1} \leftarrow J_{1,J}$ ``appear", i.e. when $J$ =2,  at at frequency of 3$(A-C)$.  These transitions fan out to higher frequencies with increasing $J$ according to the recursion formula $\nu (J) = \nu (J-1) + \frac{(B-C)}{2} J$.   The next group of transitions, for which $K_{-1}$ = 3, occur at approximately $A-\frac{(B-C)}{2}$ higher in frequency.

For the higher $K_{-1}$ members of these types of transitions, the pattern observed is such that increasingly higher $J$ transitions appear at increasingly lower frequencies as doublets, whereby the two transition types mentioned immediately above, begin to overlap one another.

\section{How Well, and Under What Restrictions, Do These Simple Formulae Work?}

The patterns, described by the formulae above, are summarized in Table 1 and shown in Figures 3 to 8.  To test the performance of these relations the following procedure was performed.  A simple FORTRAN 77 program was written (an Excel spreadsheet, or MathCad program, would achieve the same purpose) to use the relationships in Table 1 to generate the transition frequencies for a total of 75 transitions of $J$ up to 10, at different values of Ray's asymmetry parameter, $\kappa$ from -1 to -0.8.  The frequencies generated were compared to the exact frequencies for the transitions generated by a complete Hamiltonian matrix diagnolization routine using Herb Pickett's SPCAT program\cite{pick1,pick2}.  For each $\kappa$ value, and for the three cases $J'<10$ (75 transitions), $J'<6$ (43 transitions), and $J'<4$ (23 transitions), a standard deviation was calculated according to:

\begin{equation}
SD = \sqrt{ \frac{\left[ \sum (\nu (act) - \nu (est))^2 \right]}{N}}
\end{equation}

where $N$ is the number of transitions.  The results are shown in Figure 9.  It is not at all surprising that the simple formulae provided work poorly at high $J$ and high asymmetries.  If one sets a quantitative limit of desiring the predicted transitions to fall, on average, within 50 MHz of their actual location, then it is found that it is only possible to work with high $J$ transitions, i.e. $J'$ up to 9, when $\kappa$ is between -0.975 and $\approx$ -1.  If, on the other hand, it is necessary to work with an asymmetric species where $\kappa$ is -0.8, or even lower in magnitude, then it is found that the formula presented produce transition frequencies within 50 MHz of the actual frequencies only for transitions involving $J'$ less than or equal to 3.

\begin{figure}[h!]
\centering
\includegraphics[width=4.0 in]{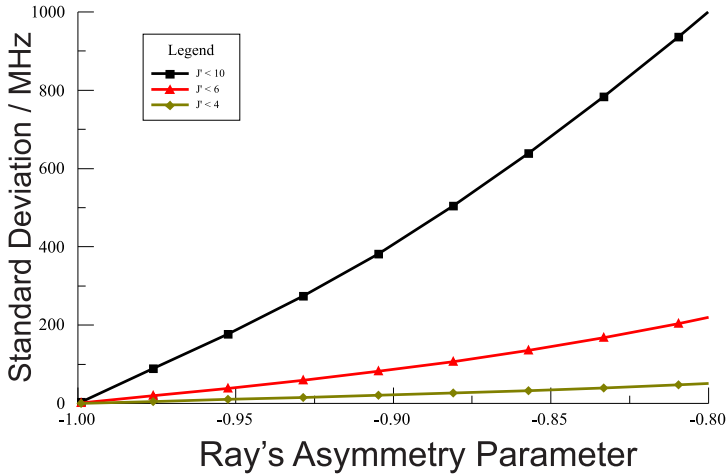}
\caption{A plot of the standard deviation in MHz pertaining to the comparision of actual rigid rotor frequencies compared to those generated using the simple formulae given in Table 1 for several different molecular asymmetries.}
\label{fig:check}
\end{figure}

This may appear very unsatisfactory.  However, the situation is better than it seems when one recalls, the very thesis of this paper, that it is the pattern of transitions that is more informative in regards to quantum number assignments than the absolute frequencies.  Figure 10 shows this comparison.  The upper portion of Figure 10 shows a rotational spectrum calculated exactly using SPCAT.  The lower portion shows a stick spectrum produced from the relationships given in Table 1 and the same rotational constants.  The match is sufficient that a quantum number assignment could likely be made for over 50 \% of the transitions shown.

\begin{figure}[h!]
\centering
\includegraphics[width=4.0 in]{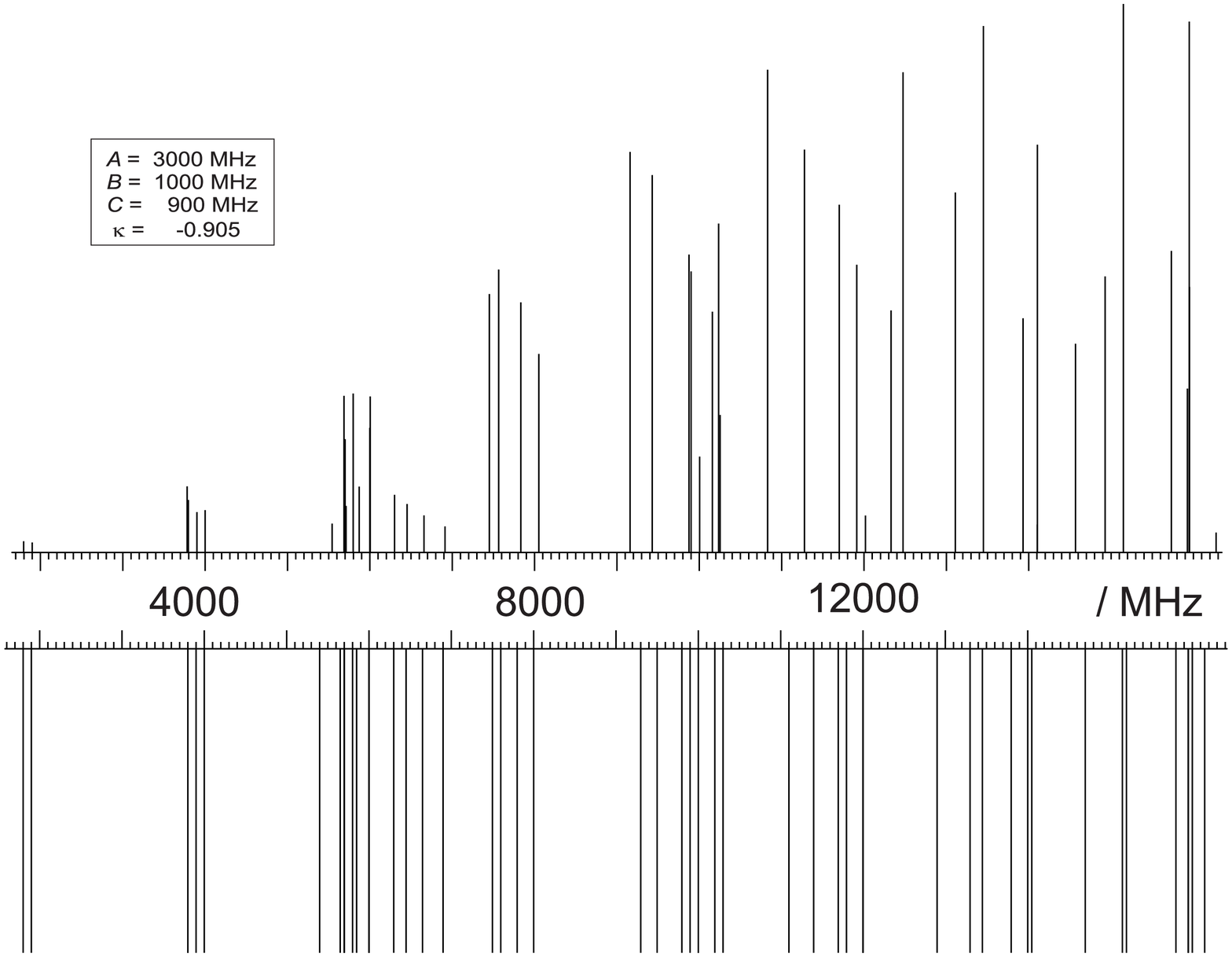}
\caption{The actual rigid rotor spectrum arising from the rotational constants given (top), compared to the rigid rotor spectrum generated using the same rotational constants and the simple formulae given in Table 1 (bottom).}
\label{fig:visual}
\end{figure}

\section{New Media Support}
In order to visualize the patterns of transitions formed by the relationships given in Table 1 both a web-based  visualization tool\cite{app} and matching smartphone mobile web page\cite{smartapp} have been created.  Using either medium, users can enter a set of rotational constants, press ``Plot Spectrum", and then see the patterns that result from the relationships given in Table 1.  The transitions displayed are colorcoded so that the $a$-type transitions are shown in red, the $b$-type transitions are in green, and the $c$-type transitions are in yellow.  A screenshot of the smartphone application is provided in Figure 11.  The display format follows closely that of the excellent AABS software package for the assignment of broadband microwave spectra developed by Prof. Z. Kisiel\cite{kis,prospe}.

\begin{figure}[h!]
\centering
\includegraphics[width=4.0 in]{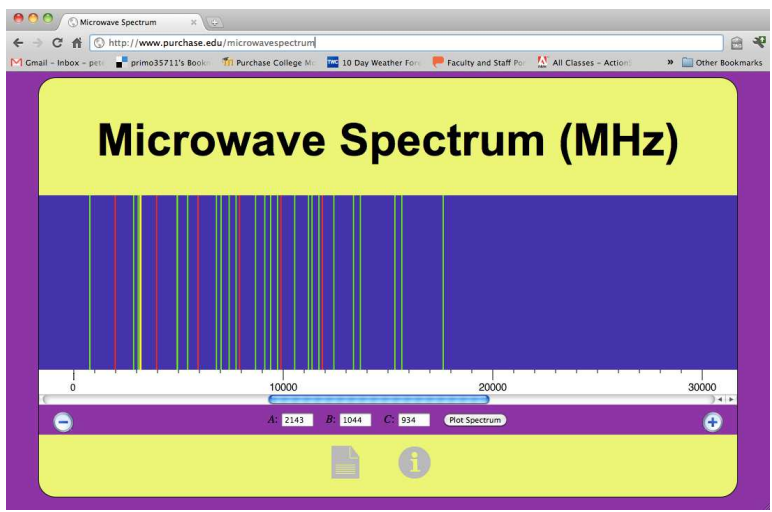}
\caption{Screenshot of the smartphone mobile website supporting this work. Users may enter values for the three rotational constants in MHz and then plot the patterns of transitions that result from the formulae of Table 1.}
\label{fig:app}
\end{figure}

\section{Conclusions}

Rotational spectra may be inspected, initially, so as to locate repeating patterns of transitions.  We have presented ``template" spectra indicating the types of patterns that can commonly occur for the near-prolate asymmetric molecule.  Location of these patterns in experimental spectra can help lead to succesful quantum number assignments and/or initial evaluations of the subject molecules rotational constants.

The information provided allows for students to adopt an alternative approach to considering the quantum mechanical, asymmetric rigid rotor.  In this method students are actively involved early, that is in the task of spectral assignment. Web and smartphone resources have been created to assist in this endeavor. The tools provided facilitate, and prompt, in depth discussion about the under lying theory and application of microwave spectroscopy. 


\section{Acknowledgement}
We are grateful to Mr. Frank DeChirico, Mr. Bryan Harrison, and Prof. S. E. Novick for proof-reading this manuscript and their improving comments.

\end{document}